\documentclass[aps,showpacs,preprintnumbers,nofootinbib]{revtex4}%
\usepackage{amssymb}
\usepackage{amsmath}
\usepackage{bm}
\usepackage{amsfonts}
\usepackage{graphicx}%
\providecommand{\U}[1]{\protect\rule{.1in}{.1in}}
\begin{document}
\title{Is Weyl \ unified theory wrong or incomplete?}
\author{Carlos Romero\thanks{E-mail address: cromero@pq.cnpq.br} \ \ }
\affiliation{Departamento de F\'{\i}sica, Universidade Federal da Para\'{\i}ba, Caixa
Postal 5008, 58059-970 Jo\~{a}o Pessoa, PB, Brazil }
\affiliation{E-mail: cromero@fisica.ufpb.br }

\begin{abstract}
In 1918, H. Weyl proposed a unified theory of gravity and electromagnetism
based on a generalization of Riemannian geometry. In spite of its elegance and
beauty, a serious objection was raised by Einstein, who argued that Weyl's
theory was not suitable as a physical theory . According to Einstein, the
theory led to the prediction of a \textquotedblleft second clock
effect\textquotedblright, which is not observed by experiments. We briefly
revisit this point and argue that a preliminary discussion on the very notion
of proper time is needed in order to consider Einstein's critical point of view.\ 

\end{abstract}

\maketitle

\vskip .5cm


\section{Introduction}

It has been widely recognized among historians and physicists that, in spite
of its elegance and beauty, the theory formulated by H. Weyl, in 1918, in his
attempt to unify gravity and electromagnetism is not suitable as a physical
theory \cite{Weyl}. As is well known, in an appendix to Weyl's paper, Einstein
set forth a serious objection to the theory. In his critique, Einstein argued
that the theory would necessarily predict the existence of the so-called
"second clock effect" \cite{Penrose}. According to Einstein, \ in a space-time
ruled by Weyl geometry the existence of sharp spectral lines in the presence
of an electromagnetic field would not be possible since atomic clocks would
depend on their past history \cite{Pauli}. Einstein reasoned that this
predicted effect is a logical consequence of Weyl's theory, insofar as in a
Weyl space-time the length of a vector is not held constant by parallel
transport, and this, in turn, would imply that the tic tac of atomic clocks,
measured by some periodic physical process, should be path dependent.

In this paper, we revisit Weyl's theory and approach Einstein's critique from
a new perspective. We argue that a preliminary discussion on the very notion
of proper time is needed in order to consider Einstein's critical point of
view.\ Our discussion will be guided by the so-called\textit{ Weyl's Principle
of Gauge Invariance,} a symmetry principle that plays an essential role in the
development of the theory.

The paper is organized as follows. In Section 2, we give a brief account of
Weyl geometry. We then proceed to Section 3 to review Einstein's argument and
examine in more detail the assumptions implicitly made therein. In Section 4,
we briefly consider recent scalar-tensor theories, which were inspired by a
weaker version of Weyl's geometry, the so-called WIST gravity theories, and
show why they are not plagued by the problem of proper time. We conclude with
some final remarks in Section 5.

\section{\bigskip A brief summary of Weyl geometry}

Weyl geometry is perhaps one of the simplest generalization of Riemannian
geometry, the only modification being the fact that the covariant derivative
of the metric tensor $g$ is not zero, but instead given by%

\begin{equation}
\nabla_{\alpha}g_{\beta\lambda}=A_{\alpha}g_{\beta\lambda},
\label{compatibility}%
\end{equation}
where $A_{\alpha}$ \ denotes the components of a one-form field $A\ $in a
local coordinate basis. This weakening of the Riemannian compatibility
condition is entirely equivalent to requiring that the length of a vector
field may change when parallel-transported \ along a curve in the
manifold\cite{Pauli}. We shall refer to the triple $(M,g,A)$\ consisting of a
differentiable manifold $M$ endowed with both a metric $g$ and a 1-form field
$A\ $as a \textit{Weyl gauge }(or,\textit{ frame)}. Now one important
discovery made by Weyl was the following. Suppose we perform the conformal
transformation
\begin{equation}
\overline{g}=e^{f}g, \label{conformal1}%
\end{equation}
where $f$ is an arbitrary scalar function defined on $M$. Then, the Weyl
compatibility condition (\ref{compatibility}) still holds provided that we let
the Weyl field $A$ transform as
\begin{equation}
\overline{A}=A+df. \label{gauge}%
\end{equation}
In other words, the Weyl compatibility condition does not change\ when we go
from one gauge $(M,g,A)$ to another gauge $(M,\overline{g},\overline{A})$ by
simultaneously transformations in $g$ and $A$.

If we assume that the Weyl connection\ $\nabla$ is symmetric,\ a
straightforward algebra shows that one can express the components of the
affine connection in an arbitrary vector basis completely in terms of the
components of $g$ and $A$:%
\begin{equation}
\Gamma_{\beta\lambda}^{\alpha}=\{_{\beta\lambda}^{\alpha}\}-\frac{1}%
{2}g^{\alpha\mu}[g_{\mu\beta}A_{\lambda}+g_{\mu\lambda}A_{\beta}%
-g_{\beta\lambda}A_{\mu}], \label{Weylconnection}%
\end{equation}
where $\{_{\beta\lambda}^{\alpha}\}$ represents the Christoffel symbols. It is
not difficult to see that the connection and, consequently, the geodesic
equations are invariant with respect to the transformations (\ref{conformal1})
and (\ref{gauge}).

We now present Weyl's second great discovery. Suppose that we are given two
vector fields $V$ and $U$ parallel transported along a curve $\alpha
=\alpha(\lambda).$ Then, (\ref{compatibility}) leads to the following
equation
\begin{equation}
\frac{d}{d\lambda}g(V,U)=A(\frac{d}{d\lambda})g(V,U).
\label{covariantderivative}%
\end{equation}
where $\frac{d}{d\lambda}$ denotes the vector tangent to $\alpha$. If we
integrate this equation along the curve $\alpha$, starting from a point
$P_{0}=\alpha(\lambda_{0}),$ we obtain \cite{Pauli}%
\begin{equation}
g(V(\lambda),U(\lambda))=g(V(\lambda_{0}),U(\lambda_{0}))e^{\int_{\lambda_{0}%
}^{\lambda}A(\frac{d}{d\rho})d\rho}. \label{integral}%
\end{equation}
Setting $U=V$ and denoting by $L(\lambda)$ the length of the vector
$V(\lambda)$ at a point\ $P=\alpha(\lambda)$\ of the curve, it is easy to
verify that in a local coordinate system $\left\{  x^{\alpha}\right\}  $ the
equation (\ref{covariantderivative}) becomes
\begin{equation}
\frac{dL}{d\lambda}=\frac{A_{\alpha}}{2}\frac{dx^{\alpha}}{d\lambda}L.
\label{length}%
\end{equation}
Let us now consider the set of all closed curves $\alpha:[a,b]\in R\rightarrow
M$, i.e, with $\alpha(a)=\alpha(b).$ Then, either from (\ref{integral}) or
(\ref{length})\ it follows that
\[
L=L_{0}e^{\frac{1}{2}%
{\displaystyle\oint}
A_{\alpha}dx^{\alpha}},
\]
where \ $L_{0}$ and $L$ denotes the values of $L(\lambda)$ at $a$ and $b$,
respectively. From Stokes's theorem we can write\footnote{Here we are assuming
that the region of integration is simply connected.}
\[
L=L_{0}e^{\frac{1}{2}\int\int F_{\mu\nu}dx^{\mu}\wedge dx^{\nu}}.
\]
We thus see that, according to the rules of Weyl geometry, the necessary and
sufficient condition for a vector to have its original length preserved after
being parallel transported along any closed trajectory is that the 2-form
$F=dA=\frac{1}{2}F_{\mu\nu}dx^{\mu}\wedge dx^{\nu}$ vanishes, where $F_{\mu
\nu}=\partial_{\mu}A_{\nu}-\partial_{\nu}A_{\mu}.$

Therefore Weyl realized that in his new geometry there are two kinds of
curvature, a \textit{direction curvature (Richtungkrummung) }and a
\textit{length curvature (Streckenkrummung}). The first is responsible for
changes in the direction of parallel transplanted vectors and is given by the
usual curvature tensor $R_{\;\beta\mu\nu}^{\alpha}$, while the other regulates
the changes in their length, and is given by $F_{\mu\nu}.$ Weyl's second great
discovery was that the 2-form $F$ is invariant under the gauge transformation
(\ref{gauge}). The analogy with the electromagnetic field is apparent and
becomes even more so when we take into account that $F$ satisfies the identity
$dF=0$ \footnote{In a local coordinate system, this identity takes the form
$\partial_{\mu}F_{\alpha\beta}+\partial_{\beta}F_{\mu\alpha}+\partial_{\alpha
}F_{\beta\mu}=0$, which looks identical to one pair of Maxwell's
equations.}$.$

\section{The principle of gauge invariance and the field equations}

Clearly, the Weyl transformations (\ref{conformal1}) and (\ref{gauge}) define
a whole equivalence class constituted by the set $\{(M,g,A)\}$ of\ all Weyl
gauges. It is then natural to expect that, as in conformal geometry, the
geometrical objects of interest are those that are gauge-invariant
\footnote{In conformal geometry, one basic invariant is the Weyl tensor
$W_{\;\beta\mu\nu}^{\alpha}$. In conformal gravity, this tensor is used to
form the scalar $W_{\;\alpha\beta\mu\nu}W_{\;}^{\alpha\beta\mu\nu}$, which,
then, defines the gravitation sector of the action \cite{Manheim}.} Surely,
these invariants will be fundamental to build the action that is expected to
give the field equations of the geometrical unified theory. Some basic
invariants are easily found: the affine connection $\Gamma_{\beta\lambda
}^{\alpha}$, the curvature tensor $R_{\;\beta\mu\nu}^{\alpha}$, the Ricci
tensor $R_{\mu\nu}=R_{\;\mu\alpha\nu}^{\alpha}$ and the length curvature
$F_{\mu\nu}=\partial_{\mu}A_{\nu}-\partial_{\nu}A_{\mu}.$ The simplest
invariant scalars, in four-dimensional space-time, that can be constructed out
of these are: $\sqrt{-g}R^{2}$,$\sqrt{-g}R_{\alpha\beta\mu\nu}R^{\alpha
\beta\mu\nu},\sqrt{-g}R_{\alpha\beta}R^{\alpha\beta}$ and $\sqrt{-g}%
F_{\alpha\beta}F^{\alpha\beta}$, where $R=g^{\alpha\beta}R_{\alpha\beta}$ is
the Ricci scalar calculated with the Weyl connection.

It seems evident that Weyl's idea was to have a physical theory completely
invariant with respect to change between gauges (or frames). Obvioulsy, this
was a minimal requirement of consistency of his physics with the new geometry.
As we know, his choice was finally to pick up the simplest possible invariant
action, namely,%
\begin{equation}
S=\int d^{4}x\sqrt{-g}[R^{2}+\omega F_{\mu\nu}F^{\mu\nu}], \label{action}%
\end{equation}
where $\omega$ is a constant \footnote{Here we are not considering the matter
action.}. Variations with respect to $A_{\mu}$ and $g_{\mu\nu}$ lead, after
some simplifications, to the field equations%
\begin{equation}
\frac{1}{\sqrt{-g}}\partial_{\nu}\left(  \sqrt{-g}F^{\mu\nu}\right)  =\frac
{3}{2}g^{\mu\nu}\left(  RA_{\nu}+\partial_{\nu}R\right)  , \label{maxwell}%
\end{equation}%
\begin{equation}
R(R_{\mu\nu}-\frac{1}{4}g_{\mu\nu}R)=\omega T_{\mu\nu}, \label{einstein-weyl}%
\end{equation}
where $T_{\mu\nu}=F_{\mu\alpha}F_{\;\nu}^{\alpha}-\frac{1}{4}g_{\mu\nu
}F_{\alpha\beta}F^{\alpha\beta}$. It is worth mentioning that when applied to
the case of a static and spherically symmetric matter distribution it can be
showed that Weyl's theory correctly predicts the perihelion precession of
Mercury as well as the gravitation deflection of light by a massive body
\cite{Pauli}. In fact, this is a consequence of the fact that all vacuum
solutions of Einstein's equations (including the Schwarzschild solution)
satisfy (\ref{maxwell}) and (\ref{einstein-weyl}) when we set $A_{\mu}=0$.

Now before we start our discussion of the Einstein's objection to Weyl's
theory, in the next section, we would like to stress that to build his theory
Weyl adopted a very strong and, at the same time, rather restrictive
principle, namely, the \textit{Principle of Gauge Invariance}, which asserts
that all physical quantities must be invariant under the gauge transformations
(\ref{conformal1}) and (\ref{gauge}). This principle was strictly followed by
Weyl and guided him \ to (\ref{action})\ when he had to choose an action for
his theory. It should also be noted here that any invariant scalar of this
geometry must necessarily be formed from both the metric $g_{\mu\nu}$ and the
Weyl gauge field $A_{\mu}.$ These two fields constitute an essential and
intrinsec part of the geometry and neither of them can be neglected when we
want to construct an invariant scalar, so they are, in this sense,
inseparable, and must always appear together.

\section{Einstein's objection revisited}

In order to examine Einstein's objection to Weyl's unified theory, let us
first spell out two of the hypotheses on which the argument is based. They can
be stated as follows:

H1) The proper time $\triangle\tau$ measured by a clock travelling along a
curve $\alpha=\alpha(\lambda)$ is given as in general relativity, namely, by
the (Riemannian) prescription
\begin{equation}
\triangle\tau=\frac{1}{c}\int\left[  g(V,V)\right]  ^{\frac{1}{2}}%
d\lambda=\frac{1}{c}\int\left[  g_{\mu\nu}V^{\mu}V^{\nu}\right]  ^{\frac{1}%
{2}}d\lambda, \label{proper time}%
\end{equation}
where $V$ denotes the vector tangent to the clock's world line and $c$ is the
speed of light. This supposition is known as the \textit{clock hypothesis }and
clearly assumes that the proper time only depends on the instantaneous speed
of the clock and \ on the metric field \cite{d'Inverno}.

H2) The fundamental tic tac of clocks (in particular, atomic clocks) is to be
associated with the (Riemannian) length $L=$ $\sqrt{g(\Upsilon,\Upsilon)}$ of
a certain vector $\Upsilon$. As a clock moves in space-time $\Upsilon$ is
parallel-transported along its worldline from a point $P_{0}$ to a point $P$,
hence $L=L_{0}e^{\frac{1}{2}\int A_{\alpha}dx^{\alpha}}$, $L_{0}$ and $L$
indicating the duration of the tic tac of the clock at $P_{0}$ and $P$, respectively.

Let us now have a look into these two assumptions. We start with the first
hypothesis (H1). First of all, we would expect that, to be consistent with the
\textit{Principle of Gauge Invariance, }proper time, as a physically-relevant
quantity, should be gauge invariant. It turns out, however, that there is no
such invariant notion of proper time in Weyl's theory. In addition to that,
the adoption of the general relativistic clock hypothesis here does not seem
to be plausible, since $\triangle\tau$, as defined above, takes into account
only part of the geometry, namely, the metric field, and completely ignores
the other geometric field, i.e., the gauge field $A_{\mu}.$ In the second
hypothesis (H2), gauge invariance is violated: the concept of tic tac is not
modelled as\ a gauge-invariant physical quantity and, again, the Weyl
geometrical field plays no role in its determination.

To conclude this section, let us remark that, with the inexistence of an
invariant notion of proper time, even the \textit{first clock effect }(the
"twin paradox"), which appears both in the special and general relativity,
cannot be predicted in Weyl's theory.

\section{The incompleteness of Weyl's theory}

Since it does not come equipped with an appropriate notion of proper time,
consistent with the requirement of gauge invariance, we are \ forcebly led to
conclude Weyl's theory is not complete. An interesting question that now
arises is whether or not one could come up with an acceptable definition of
proper time ($\triangle\tau)$ in this theory. Of course the sought-after
definition would have to fulfill the following requirements:

i) $\triangle\tau$ should be constructed entirely from the geometry (recall
that the general relativistic proper time (\ref{proper time}) is
proporportional to length);

ii) $\triangle\tau$ should be consistent with the Principle of Gauge Invariance;

iii) $\triangle\tau$ should depend both on the metric field $g_{\mu\nu}$ and
the gauge field $A_{\mu}$ \footnote{In order to preserve gauge invariance it
is expected that $A_{\mu}$ should appear in $\mathcal{F}$ only via the tensor
$F_{\mu\nu}.$};

iv) $\triangle\tau$ should reduce to the general relativistic definition of
proper time (\ref{proper time}) in the limit when $A_{\mu}$ goes to zero.

v) $\triangle\tau$ should be written in the form $\triangle\tau=\int
\mathcal{F}(V,g,A)d\lambda,$ with $\mathcal{F}$, as in Finsler geometry, being
a first-order homogeneous function with respect the tangent vector $V$ (This
condition is necessary to garantee invariance under reparametrization)
\cite{Finsler}.

vi) Finally, it would be highly desirable, though not strictly necessary, that
the new definition of \ $\triangle\tau$ could allow for the Weyl affine
geodesic equations to be deduced from a variational principle \footnote{At
present, we do not know if this requirement is mathematically possible. In
fact, the solution of this question will lead us to examine the
so-called\textit{ inverse variational problem }for the case of Weyll affine
geodesics\textit{ }\cite{Shabanov}.}.

To find a good definition of proper time that fulfills all the above
requirements does not seem to be an easy task. First of all, because of the
condition on homogeneity with respect to $V$ one has to look for second-order
gauge-invariant tensors. Candidates that immediately come to mind are:
$R_{\mu\nu}$and $Rg_{\mu\nu}.$ These would lead, respectively, to
$\triangle\tau_{1}=a_{1}\int\left[  R_{\mu\nu}V^{\mu}V^{\nu}\right]
^{\frac{1}{2}}d\lambda$ and $\triangle\tau_{2}=a_{2}\int\left[  Rg_{\mu\nu
}V^{\mu}V^{\nu}\right]  ^{\frac{1}{2}}d\lambda$, $a_{1}$ and $a_{2}$ denoting
dimensional constants. However, it is clear that neither $\triangle\tau_{1}$
nor $\triangle\tau_{2}$ is a good choice as they do not satisfy conditions
(iv) and (vi). A rather contrived choice would be $\triangle\tau_{3}=a_{3}%
\int\left[  \frac{g_{\mu\nu}V^{\mu}V^{\nu}}{g_{\mu\nu}W^{\mu}W^{\nu}}\right]
^{\frac{1}{2}}d\lambda$, where $a_{3}$ designates a dimensional constant and
$W$ is the gauge-invariant 1-form defined by $W=A+d(\ln(R))$, with $W^{\mu
}=g^{\mu\nu}W_{\nu}.$ In this case, it is interesting to note that the light
cone structure is gauge invariant and is determined by the metric only.
However, again this choice is not consistent with conditions (iv) and (vi).
Perhaps the solution of the problem of finding a satisfactory definition of
proper time may lead us beyond the Weyl geometrical framework, indicating that
we need a higher level of generalization, such as the one we find in Finsler
geometry \cite{Finsler1}.

\section{Final remarks}

When $A$ is an exact form, i.e., $A$ $=d\phi,$ where $\phi$ is a scalar field,
then we say that we have an \textit{integrable Weyl space-time} (WIST).
Theories framed in this kind of geometry are not subject to Einstein's
objection since they do not predict a second clock effect. Due to this fact
WIST gravity has attracted the attention of some cosmologists \cite{Novello}.
Instead of the electromagnetic field, now it is the scalar field that is
geometrized. It is interesting to recall here that in the case of WIST
theories the definition of proper time is given by the gauge-invariant
equation \cite{Romero}
\begin{equation}
\Delta\tau=\int_{a}^{b}e^{-\frac{\phi}{2}}\left(  g_{\mu\nu}\frac{dx^{\mu}%
}{d\lambda}\frac{dx^{\nu}}{d\lambda}\right)  ^{\frac{1}{2}}d\lambda.
\label{proper time wist}%
\end{equation}
It is not difficult to verify that, with this definition, $\Delta\tau$
satisfies all the requirements listed in the previous section, with $\phi$
replacing $A_{\mu}$.

We do not know if it is possible to "complete" the elegant and profound theory
developed by Weyl almost a century ago \footnote{Pauli refers to Weyl's theory
as "an extremely profound theory" \cite{Pauli}.}. Perhaps in a modified
version, but still inspired in Weyl's ideas, the essential features of the
theory could be revived. As some authors have put it: "Weyl geometrical theory
contains a suggestive formalism and may still have the germs of a future
fruitful theory "\cite{Bazin}.

\bigskip

\section*{Acknowledgements}

\noindent The author would like to thank J. B. Fonseca-Neto, F. Dahia, R. G.
Lima and J. B. Formiga for helpful discussions. This work was partially
supported by CNPq (Brazil).

\end{document}